\documentclass[aps,twocolumn]{revtex4}
\usepackage{graphics}
\usepackage{epsfig}
\usepackage{color}

\begin{document}
\title{Kolmogorov scaling from random force fields}
\author{Mogens H.~Jensen $^1$, Kim Sneppen $^1$ and Luiza Angheluta $^2$}

\email{mhjensen@nbi.dk,sneppen@nbi.dk,luiza.angheluta@fys.uio.no}
\affiliation{$^1$ Niels Bohr Institute, Blegdamsvej 17, Dk 2100,
Copenhagen, Denmark\\
$^2$ Center for Physics of Geological Processes, Univ. of Oslo,
Oslo, Norway} \homepage{http://cmol.nbi.dk}
\date{\today}

\begin{abstract}
We show that the classical Kolmogorov and Richardson scaling laws in
fully developed turbulence are consistent with a 
random Gaussian force field.
Numerical simulations of a shell model approximation to 
the Navier-Stokes equations suggest that the
fluctuations in the force (acceleration) field are scale independent
throughout the inertial regime. We conjecture that Lagrangian
statistics of the relative velocity in a turbulent flow is
determined by the typical force field, whereas the multiscaling is
associated to extreme events in the force field fluctuations.
\end{abstract}

\maketitle

In studies of fully developed turbulence, two discoveries are highly
noticeable as fundamental and seminal. One regards L.R. Richardson's
study of the enhanced dispersion of particles advected by a
turbulent flow~\cite{Richardson26}. The other result is Kolmogorov's
fundamental derivation, essentially based on dimensional arguments,
of the energy spectrum in fully developed
turbulence~\cite{Kolmogorov41}. Both theories employ the energy
cascade, from the integral scale down to the dissipation scale, as
the paradigmatic physical picture of the energy dissipation flow.
Indeed, pair-particles passively advected by turbulence exhibit a
superdiffusive behavior with their relative distance given by the
Richardson's scaling ~\cite{Boffetta02},~\cite{Boffetta02bis}. In
contrast to the pair dispersion, the motion of a single particle is
determined by the correlation time of the underlying velocity field,
such that it is transported ballistically for times smaller than the
correlation time and diffuses normally for larger time
scales~\cite{Falkovich01}. In the cascade scenario, the
pair-particle superdiffusion is due to the large energy jumps
between the eddies containing each particle. However, in the
velocity space no superdiffusive behavior is needed to substantiate
this jumpy motion.

In this Letter, we show that the velocity increments generated by a
white-noise force field are sufficient to generate the
superdiffusive behavior, as well as the Kolmogorov energy spectrum.
To put it in very simple terms: integrating 'up' from the random
acceleration field to the velocity field and subsequently to the
displacement is enough to reproduce the well-known scaling laws.

\begin{figure}[]
\centerline{\epsfig{file=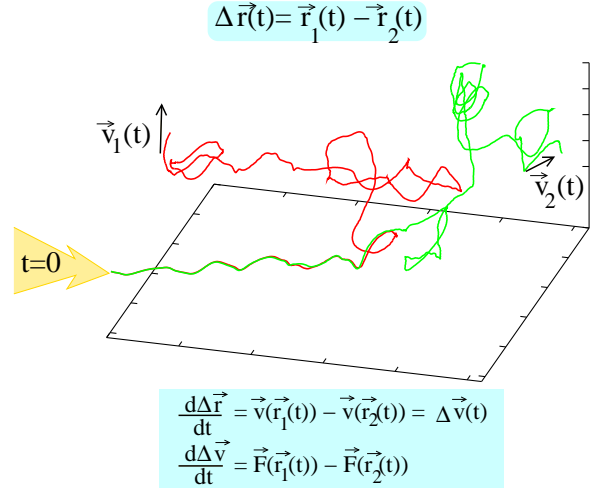,angle=0,width=8.5cm,clip=}}
\caption{\small Two particles being advected in a random force
field, generated by the GOY shell model.
\label{goy}}
\end{figure}

To clarify the underlying physical picture, we consider a simple
stochastic model of relative dispersion in a white noise
acceleration field given by
\begin{eqnarray}
\frac{d\Delta v}{dt} & = & \Delta F(t)\\
\langle \Delta F(t')\Delta F(t") \rangle & = & 4 \epsilon^* \;\delta
(t'-t"), \label{epsilon}
\end{eqnarray}
where $\Delta v(t) = v_1(t)- v_2(t)$ is the velocity difference
between the two particles moving along the two trajectories $r_1(t)$
and $r_2(t)$, and $\Delta F(t) = F_1(t)-F_2(t)$ is the relative
force. The prefactor $4$ instead of the usual factor $2$ appearing
in the force correlation is due to the parametrization of the
relative dispersion in terms of the diffusion constant $\epsilon^*$
for a single particle in the velocity space. The $\delta-$function
may have, in principle, a finite width given by the time correlation
of the relative random field along the two trajectories. This width
will be determined both by the time it takes to pass a correlation
length for a given force realization, and the time it takes to
change the force in a certain point of the system.

In this set up, the relative velocity field, $\Delta v(t) = \int_0^t
\Delta F(s)ds$, is a Wiener process with a Gaussian distribution,
namely
\begin{equation}\label{vcorrel}
\langle \Delta v(t_1) \Delta v(t_2)\rangle = 4\epsilon^* \;
min(t_1,t_2),
\end{equation}
while the relative separation is described by a non-Gaussian
distribution with the second moment satisfying the Richardson's
scaling, that is
\begin{equation}
\langle \Delta r^2(t)\rangle = \frac{4\epsilon^*}{3}t^3\label{rdiff}.
\end{equation}
By eliminating the time dependence of the relative velocity and
distance, we obtain the exact Kolmogorov scaling,
\begin{equation}\label{kolmog}
\langle \Delta v^2(t)\rangle = 48^{1/3}\epsilon^{*2/3}\langle\Delta
r^2(t)\rangle^{1/3},
\end{equation}
in the Lagrangian framework (for the higher moments see~
\cite{moments}). Thus, Kolmogorov scaling is consistent with the
assumption that the dispersion is driven by sufficiently random and
uncorrelated acceleration fields.

In deriving Eq.~(\ref{kolmog}) we assumed that the relative velocity
is obtained by following the Lagrangian trajectories, which in a
real turbulent flow may differ from the typical velocity increments
separated by the distance $r$ (Eulerian measurement of the velocity
differences)~\cite{Mordant03}.

\begin{figure}[]
\epsfig{file=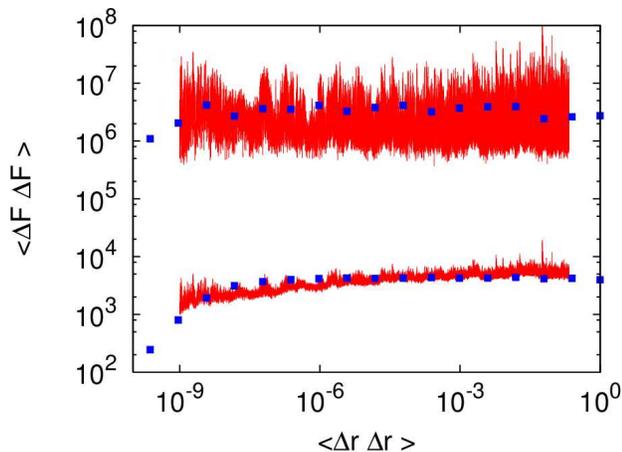,height=8.5cm,angle=270,clip=}
\caption{\small The squared relative acceleration $\Delta F \Delta F$
and its infinite moment versus $\Delta r^2$. The thin lines
are for the Lagrangian trajectories where distances and accelerations are
parameterized by the time of advection. The squares represent the
corresponding Eulerian measures of the same quantities.
}
\label{accel}
\end{figure}

Eq.~\ref{epsilon} implies that $2 \epsilon^*$ is the diffusion
constant for the relative velocity.  For a Lagrangian stochastic
flow generated by the white noise acceleration field,  $\epsilon^*$
can be estimated as
\begin{eqnarray}
\nonumber \langle \Delta \vec v(t) \Delta \vec F(t) \rangle & = &
\langle \int_0^t ds \Delta \vec F(s)\cdot \Delta \vec F(t)\rangle\\
& = & \nonumber \int_0^t ds \langle \Delta \vec F(s)\cdot \Delta \vec F(t)\rangle\\
& = & 12 \epsilon \int_0^t \delta(t-s) ds = 12 \epsilon^*,
\label{epsilon-est}
\end{eqnarray}
where the additional factor 3 compared to eq.~\ref{epsilon} is
related to the 3D system. From dimensional considerations,
$\epsilon^*$ has the same units [$length^2/time^3$] as the standard
energy dissipation rate $\epsilon$ characterizing the turbulence
cascade.

To examine how the Lagrangian white-noise acceleration relates to
the anomalous scaling laws in a more realistic turbulent field, we
consider the kinematics of pair particles advected by the
homogeneous turbulent flow obtained by a real-space transformation
of the GOY shell model~\cite{Jensen99,Bohr98}. This model proposed
originally by Glenzer, Yamada and Ohkitani~\cite
{Gledzer73,Yamada87} provides a description of the turbulent motion
embodied in the Navier-Stokes equations. The GOY model is formulated
on a $N$-discrete set of wavenumbers, $k_n=2^n$, with the associated
Fourier modes $u_n$ evolving according to
\begin{eqnarray}
\label{un}
(\frac{d}{ dt}+\nu k_n^2 ) \ u_n \ & = &
 i \,k_n (a_n \,   u^*_{n+1} u^*_{n+2} \, + \, \frac{b_n}{2}
u^*_{n-1} u^*_{n+1} \, + \, \nonumber \\
& & \frac{c_n }{4} \,   u^*_{n-1} u^*_{n-2})  \ + \ f \delta_{n,1},
\end{eqnarray}
for $n=1\cdots N$. The coefficients of the non-linear terms are
constrained by two conservation laws, namely the total energy, $E =
\sum_n |u_n|^2$, and the helicity (for 3d), $H =
\sum_n(-1)^{n}k_n|u_n|$, or the enstrophy (for 2d), $Z = \sum_n
k_n^2|u_n|^2$, in the inviscid limit, i.e.
$f=\nu=0$~\cite{Biferale95}. Therefore, they may be expressed in
terms of a free parameter only $\delta\in [0,2]$, $ a_n=1,\
b_{n+1}=-\delta,\ c_{n+2}=-(1-\delta)$. As observed by
Kadanoff~\cite{Kadanoff95}, one obtains the canonical value
$\epsilon= 1/2$, when the 3d-helicity is conserved. The set
(\ref{un}) of $N$-coupled ordinary differential equations can be
numerically integrated by standard techniques~\cite{Pisarenko93}. We
have used standard parameters in this paper $N = 19, \nu = 10^{-6},
k_0 = 2 \cdot 10^{-4}, f = 5 \cdot 10^{-3}$.

The GOY model is defined in $k$-space but we study particle
dispersion in direct space obtained by an inverse Fourier
transform~\cite{Jensen99} of the form
\begin{equation}
\label{field}
 {\vec v} ({\vec r},t)=\sum_{n=1}^{N}
 {\vec c}_n [u_n(t) e^{i {\vec k}_n\cdot {\vec r}}
+ c.\, c.].
\end{equation}
Here the wavevectors are ${\vec k}_n ~=~ k_n {\vec e}_n$ where
${\vec e}_n$ is a unit vector in a random direction, for each shell
$n$ and ${\vec c}_n$ are unit vectors in random directions. We
ensure that the velocity field is incompressible, $\nabla\cdot{\vec
v} =0$, by constraining ${\vec c}_n \cdot {\vec e}_n =0,~\forall n
$. In our numerical computations we consider the vectors ${\vec
c}_n$ and ${\vec e}_n$ quenched in time but averaged over many
different realizations of these.

As an example of the motion in this field, Fig.~\ref{goy} shows the
trajectories of two passively advected particles. As the relative
distance diverges in time, the two particles experience different
force fields, which in turn typically increase the difference in the
relative velocities of the two particles.  The figure shows the
individual particles as they are advected, first together and later
diverging away from each other when they are encased in different
eddies.

Fig.~\ref{accel} examines the noise in the effective force field
$\langle \Delta F\Delta F\rangle$ for the relative motion of the two
advected particles. In Fig~\ref{accel}a) we use viscosity
$\nu=10^{-6}$, with a Kolmogorov scale $\Delta r \sim 1.0\cdot
10^{-4}$. The noise amplitude is plotted versus the average square
distance between the particles $\langle \Delta r^2\rangle=\langle(
{\bf r}_1(t)-{\bf r}_2(t))^2\rangle$ , with the time as
parametrization of the curves, as in eq. \ref{kolmog}. The average
is over independent trials of the two advected particles. One
observes that both the typical noise and the maximum value at any
distance is constant throughout the inertial range, i.e. above the
Kolmogorov scale.

\begin{figure}[]
\epsfig{file=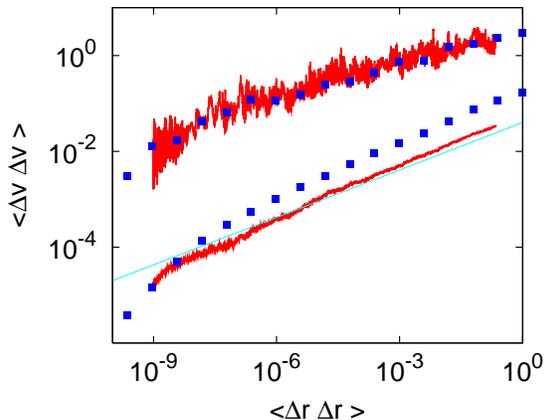,width=8.5cm,angle=0,clip=}
\caption{\small The squared relative velocity $(\Delta v)^2$
and its infinite moment versus $(\Delta r)^2$ (ie a representation
of the structure function). The thin lines
are for the Lagrangian trajectories where distances and velocities are
parametriced by the time of advection. The filled squares represent are the
corresponding Eulerian measures. The straight line represents 
standard Kolmogorov scaling 
$\langle \Delta v^2 \rangle \propto \langle \Delta r^2\rangle^{1/3}$.
}
\label{structure}
\end{figure}

We conclude that the force field is equivalent to Gaussian white
noise, and therefore the structure function of this turbulent field
should be close to the one predicted by Eq. (\ref{kolmog}). This is
confirmed in Fig. \ref{structure} where we show the deviations in
velocity as a function of the square distance between the particles.
One indeed sees that $\langle \Delta v^2(t) \rangle$ versus $\langle
\Delta r^2(t)\rangle$ scales with an exponent close to $1/3$ in
agreement with our expectations. For completeness, we also show the
infinite moment of the velocity, and remark that this higher moment
scales with an exponent close to $0.23$. This signals
multidiffusion~\cite{Sneppen94} where extreme velocity differences
sometimes, but rarely, are reached after short separations. In the
current context, we see these extreme deviations as a measure of
very unlikely and intermittent events which only add little to the
typical behavior of the flow. Indeed also the Eulerian statistics
shows clear multiscaling as expected \cite{Jensen99}.

While our intuition has been based on the Lagrangian picture of
advected particles, it is remarkable that the corresponding Eulerian
quantities behaves similarly. This is demonstrated in simulations
where we now fix the distance between two points, and then calculate
respectively the difference in velocity and acceleration. The
continuous curves in Figs. \ref{accel} and \ref{structure} show how
$\langle \Delta F^2(r)  \rangle$ and $\langle \Delta v^2(r) \rangle$
vary with the square relative distance between the investigated
points. From Fig. \ref{accel} we see that the value of the plateau
for the random force field is a direct consequence of its random
expectation at any large distance. Therefore, there is nothing
special about the selection of advected points in the Lagrangian
case. In fact the onset of the plateau is slightly sharper in the
Eulerian case, presumably reflecting averaging associated to the
underlying time parameter in the Lagrangian advection. Similarly,
there is no significant difference for the structure functions shown
in Fig. \ref{structure}.

\begin{figure}[]
\epsfig{file=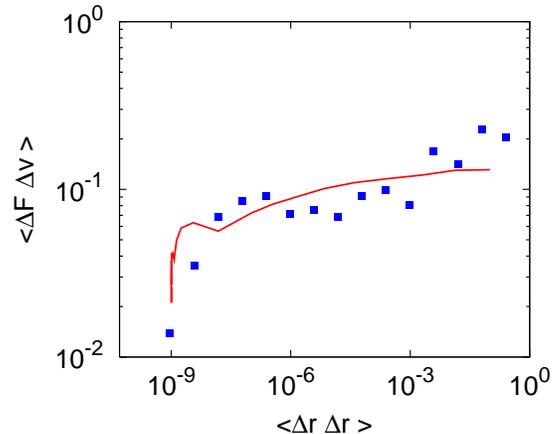,width=8.5cm,angle=0,clip=}
\caption{\small A measure of an effective diffusion
constant in velocity: $\langle \Delta v \Delta F \rangle$
versus $\Delta r^2$. As in the previous figures the thin line
describes Lagrangian trajectories 
parametriced by the time of advection. The filled squares  
correspond to the Eulerian case.
}
\label{epsilon}
\end{figure}

Using Eq. \ref{epsilon-est} we estimate $\langle \Delta v \Delta F
\rangle \sim 0.1$ throughout the inertial range in the GOY model
simulations, see Fig. \ref{epsilon}. This value of the effective
velocity diffusion constant is larger than the average energy
dissipation at the Kolmogorov scale, estimated from the energy input
${\rm Re} \langle u_1^* \cdot f\rangle=0.001$ in the GOY model. This
discrepancy in the effective diffusion terms we attribute to the
huge contributions from the spikes in the acceleration which are
absent in the simple white noise calculation of Eq.
\ref{epsilon-est}. These spikes also gives rise to multidiffusion,
as discussed above.

We believe that the acceleration field as shown in Fig. \ref{accel}
should be experimentally accessible either by particle tracking in a
3-D flow~\cite{Luthi07} or from probe measurements in channel flows
employing the Taylor hypothesis. In the first case the acceleration
is easily estimated from the temporal variations in the velocity
field of the 3-D advected particles.

Overall we have seen that the force field reaches an average value
that is independent on the distance between the advected points in
the turbulent fluid. Already at distances slightly above the
Kolmogorov scale the two particles often receive random ``kicks"
which are as large at small scales as they are at the integral
scale. Thus, huge accelerations are associated to the very small
scales, presumably to the core of eddies at the verge of their
destruction by dissipation. The acceleration between two particles
are primarily dependent on how close each of them are to the center
of an eddy. 
When examining the distribution of the accelerations at a fixed distance
we observe a broad power law like behavior with a cutoff
which is independent of the distance (as demonstrated by the
constant max norm). The size of the cutoff is determined by the size
of the forcing and the scale at which this forcing is acting (in our
simulations, the scale is $\Delta r=1$).

In conclusion, the motion associated to the relatively slow turn
over dynamics of the large eddies is not needed for obtaining
Richardson or Kolmogorov statistics. These two seminal laws are
primarily a consequence of the random force field that fluctuates
with an amplitude set by the system size and with a correlation time
set by the Kolmogorov scale.
\\

We are grateful to Hiizu Nakanishi, Simone Pigolotti and Yves
Pomeau for valuable discussions.
We thank the Danish National Research Foundation for support through
the Center for Models of Life.


\begin{thebibliography}{}
\bibitem{moments}
Using Wick's theorem we obtain the $2n$-th moment of the velocity
difference at time $t$: $\langle \Delta v^{2n}(t)\rangle =
\frac{(2n)!}{2^n n!} (48\epsilon^{*2})^{n/3}\langle\Delta
r^2(t)\rangle^{n/3}$
\end{thebibliography}
\end{document}